\documentclass[12pt,preprint]{aastex}
\begin{document}

\newcommand\grbs{$\gamma$-ray bursts}

\title{Pre-Existing Superbubbles as the Sites of Gamma-Ray
Bursts}
\author{John Scalo$^1$ \& J. Craig Wheeler$^1$\\
}
\affil{$^1$Astronomy Department, University of Texas, Austin, Texas
78712;\\
parrot@astro.as.utexas.edu,
 wheel@astro.as.utexas.edu}

\begin{abstract}

According to recent models, $\gamma$-ray bursts apparently
explode in a wide variety of ambient densities ranging from
$\sim 10^{-3}$ to 30 cm$^{-3}$. The lowest density environments
seem, at first sight, to be incompatible with bursts in or near
molecular clouds or with dense stellar winds and hence with
the association of $\gamma$-ray bursts with massive stars.  We 
argue that low ambient density regions naturally exist in areas
of active star formation as the interiors of superbubbles.
The evolution of the interior bubble density as a function of time
for different assumptions about the evaporative or hydrodynamical
mass loading of the bubble interior is discussed.
We present a number of reasons why there should exist a
large range of inferred afterglow ambient densities whether
$\gamma$-ray bursts arise in massive stars or some version of
compact star coalescence.   We predict that many
$\gamma$-ray bursts will be identified with X-ray bright
regions of galaxies, corresponding to superbubbles, rather than
with blue localized regions of star formation.
Massive star progenitors are expected to have their own circumstellar winds.  
The lack of evidence for individual stellar winds associated with the 
progenitor stars for the cases with afterglows in especially low density 
environments may imply low wind densities and hence low mass loss rates 
combined with high velocities.  If $\gamma$-ray bursts are
associated with massive stars, this  combination might be
expected for compact progenitors with atmospheres  dominated by
carbon, oxygen or heavier elements, that is, progenitors 
resembling Type Ic supernovae.

\end{abstract}


\newpage

\section{INTRODUCTION}

There is circumstantial evidence that $\gamma$-ray bursts are
associated with the collapse of massive stars.  The events seem
to occur in galaxies with active star formation (Hogg \&
Fruchter 1999).  Sokolov et al. (2001) use spectral synthesis of
$\gamma$-ray burst host galaxies  to conclude that their sample
galaxies have large SFRs and only appear to be below L$_*$
because of dust extinction.  Some $\gamma$-ray burst afterglows
seem to reveal evidence for supernova light (Bloom et al. 1999;
Reichart, 1999; Galama et al. 2000).   Recent observations
reveal evidence for iron that may be ejected from the explosion
(Antonelli et al. 2000; but note that the abundance of the iron
is very model dependent and that the observed features may be
consistent with a solar abundance of iron, Rees \&
M\'esz\'aros  2000).

If $\gamma$-ray bursts occur in massive stars, then there are
two expectations for their environment.  The immediate
environment should be dominated by a strong stellar wind and
the larger environment should be typical of the star-forming
region, with densities of at least, and perhaps significantly
exceeding, $10-100$ cm$^{-3}$, characteristic of large
molecular clouds complexes. By contrast, recent multiwavelength
analysis of selected
$\gamma$-ray burst afterglows by Panaitescu \& Kumar (2001)
have shown that the ambient density of some $\gamma$-ray bursts
can be as low as
$\sim 10^{-1}$ to $10^{-3}$ cm$^{-3}$, and perhaps even less,
for their particular afterglow shock model.  These low densities
seem to be incompatible with the hypothesis that
$\gamma$-ray bursts are associated with massive stars.  We
argue here that, on the contrary, such small densities can be
understood if
$\gamma$-ray bursts explode within pre-existing interiors of
superbubbles, themselves the remnants of  earlier massive star
formation, and that even the range in densities can be
understood.  We investigate the implications of this hypothesis
for the nature of $\gamma$-ray bursts.

In \S 2 we summarize the information on the ambient densities of
$\gamma$-ray bursts and describe models of superbubbles and
their evolution. In \S3 we discuss the expected variation in
afterglow densities and in \S4 we outline various ways in which
$\gamma$-ray bursts could be born in and interact with
superbubbles.  Our conclusions, including important constraints
on the progenitor wind, are presented in \S 5.

\section{AFTERGLOWS AND SUPERBUBBLES}

Panaitescu \& Kumar (2001) analyze the multi-wavelength data of
the afterglows of four well-studied $\gamma$-ray bursts,
assuming that the emission is due to the interaction of a
collimated relativistic shock with the ambient medium, and
subsequent emission of synchrotron and inverse
Compton-scattered radiation. They find that each of these
bursts is incompatible with the interaction with a 1/r$^2$ wind
but is compatible with an interstellar medium of constant
density.  The values they derive for the ambient density are
remarkably low: GRB 980703
$\sim 8.0\times10^{-4}$ cm$^{-3}$; GRB 990123 $\sim
8.1\times10^{-3}$ cm$^{-3}$; GRB 990510
$\sim 2.2\times10^{-1}$ cm$^{-3}$; GRB 991216
$\sim 2.4\times10^{-4}$ cm$^{-3}$.
Other studies have obtained a range, generally higher, for the
ambient density.  Wijers et al. (1999) and Frail et al. (2000) find
about 0.5 cm$^{-3}$ for GRB 970508.
Higher densities, $\sim$ 30 cm$^{-3}$, have been associated
with some events (Kumar, private communication, 2001; Harrison et al. 2001),
and Piro et al. ascribe a density of $\approx 4\times 10^4$ cm$^{-3}$
to GRB 000926.

At face value, the lowest of these ambient densities and the
lack of evidence for a 1/r$^2$ stellar wind contradict the hypothesis
that the ``long" $\gamma$-ray bursts with afterglows arise in
massive stars. Massive stars must inevitably blow a stellar
wind and they are often associated with dense interstellar
clouds.  Both of these issues must be addressed if the
association of $\gamma$-ray bursts with massive stars is to be
maintained in the presence of low ambient densities. Here we
focus on the properties of superbubbles, but the constraint of
the wind remains severe.  We return to that topic in the
discussion of
\S 5.  

The low ambient density for some $\gamma$-ray bursts is
actually not so exotic, but is characteristic of the densities
inside ``superbubbles" formed by the H II regions, winds, and
supernovae of clusters of massive stars (Weaver et al. 1977;
Tomisaka \& Ikeuchi 1986; McCray \& Kafatos 1987; MacLow \&
McCray 1988), just the environment one might associate with
massive star prognitors of $\gamma$-ray bursts.  There is
observational evidence for such structures and associated
low-density interiors.  The ``Local Hot Bubble" and the Loop I
Superbubble in Sco-Oph, which are currently interacting, have
interior densities estimated from model fits to X-ray spectra
of about
$2 \times 10^{-3}$ cm $^{-3}$ for the Local Bubble and
$2-5\times 10^{-2}$ cm $^{-3}$ for the Loop I Superbubble (see
Breitschwerdt, Freyberg, \& Egger 2000 and references therein).
Many external galaxies show evidence for large HI holes that may
be associated with superbubbles (see Walter 1999 for a
summary).

The density in a superbubble depends on a number of parameters,
the ambient density into which the bubble expands (itself perhaps
porous), the time-dependent power input from H II regions,
winds, and supernovae, the evaporation of clouds and of the compressed
shell of ambient gas, turbulent attrition of the shell,  and the time since
the onset of the power
input, among others.  To represent the density evolution, we adopt
the expression from Shull \& Saken (1995) for the interior density
of the bubble.  Shull \& Saken assume an isothermal interior in
pressure balance and hence derive an interior density
loaded by mass evaporation from the shell that is
radially constant interior to the bubble shell.  They give
(their Eq. 12):
\begin{equation}
{\rm n_b(t) = 1.6\times10^{-2}~{\rm cm^{-3}}~L_{38}^{6/35}n_a^{19/35}
t_6^{-22/35}k_o^{2/7}},
\end{equation}
where L$_{38}$ is the power input from winds and supernovae in
units of $10^{38}$ erg~s$^{-1}$, n$_a$ is the ambient number
density into which the bubble propagates, t$_6$ is the time
since the bubble was initiated in units of $10^6$ yrs, and
k$_0$ is a factor of order unity that accounts for possible
suppression of conductivity by magnetic fields or enhancement
by evaporation of engulfed clouds (Silich et al. 1994, 1996). 
Eqn. (1) is very similar to the formula given
by MacLow \& McCray (1988) based on the solution of Weaver et
al. (1977), except that MacLow \& McCray include a spatial
dependence factor
$(1-r/R)^{-0.4}$, where R is the shell radius, that
causes the density to rise near the shell.

    The numerical coefficient in Eqn. (1),
and perhaps even the scaling with
parameters, depends on the assumption that the bubble
interior is mass-loaded by classical evaporation from the interior of the
shell.  Silich et al. (1994, 1996) have performed
three-dimensional non-hydrodynamic simulations of superbubbles expanding
into cloudy ambient media with different cloud filling
factors (and other parameters) and find that the mass loading is dominated
by evaporating engulfed clouds rather than evaporation
from the superbubble shell.  Figure 1 in Silich et al. (1994) indicates
interior densities larger than given by Eqn. (1) (when
scaled
to the same L$_{38}$) by a factor of 3 at 10 Myr for an assumed cloud
filling
factor of 0.1.  Inspection of their figures indicates
that the time scaling of n$_b$(t) is roughly consistent with the t$^{-2/3}$
scaling in Eqn. (1), although it depends somewhat on the
parameters.  
    
In contrast, magnetic fields can suppress conduction even if they are
dynamically unimportant.  Strickland \& Stevens (1998) present
two-dimensional hydrodynamic simulations of wind-blown bubbles in which
evaporation is completely neglected.  In their
simulations, the interior bubble density is determined by the mixing of
dense shell material into the hot interior due to shear
motions between the interior and the dense shell that is corrugated by
instabilities.  Their Figure 4 shows an order of magnitude
decrease of the interior bubble density relative to the Weaver et al. (1977)
classical conduction solution.  Strickland \& Stevens
do not present the time evolution, so we cannot say whether the time scaling
would be similar to Eqn. (1).

        That the time dependence of n$_b$ in Eqn. (1)
 depends on the
type of mass loading of the bubble interior can be seen by
considering the rate of change of n$_b$ due to mass loading at
rate $\rm{\dot M(t)}$ and
bubble expansion as
\begin{equation}
{\rm \frac{dn_b}{dt} = \frac{3{\dot M}}{4\pi R^3 m_p} - 3 \frac{n_b}{R}
\frac{dR}{dt}},
\end{equation}
where $\rm{m_p}$ is the average mass of a particle in the bubble interior.
If the shell radius scales as R $\propto$ t$^{3/5}$ (Weaver et al. 1977;
this implies n$_a$ = const, see below), then the second
term (no mass loading) gives a contribution to n$_b$(t) $\propto {\rm
R^{-3}}$ 
that varies as t$^{-9/5}$.
For the first (mass loading) term, Shull \& Saken find a classical
conduction mass input rate ${\dot M}$ that scales approximately as
t$^{1/6}$.  Using R $\propto$ t$^{3/5}$, this term gives a contribution
to n$_b$(t) that scales as t$^{-19/30}$,
just the scaling (t$^{-2/3}$) given by
Shull \& Saken, showing that their result for n$_b$(t) is dominated
by the conductive mass loading.  Since
mass loading by hydrodynamic effects should occur even in the absence
of conduction (Strickland \& Stevens 1998), the no-conduction case
should lie in between these two extremes.  For example, if ${\dot M}$
and n$_a$ are constant, then the mass loading term yields n$_b$(t) $\propto$
t$^{-4/5}$.  We continue to use the classical conduction
solution t$^{-2/3}$ as given by Shull \& Saken,
with the understanding that the time dependence may be steeper.
In what follows, we will adopt the approximate expression for the
bubble interior density to be:
\begin{equation}
{\rm n_b(t) = 1.6\times10^{-2}~{\rm cm^{-3}}~L_{38}^{1/6}n_a^{4/7}
t_6^{-2/3}\theta},
\end{equation}
where $\theta$ accounts for the uncertainty in the mass loading
rate.  The discussion above suggests $0.1 \la \theta \la 3$.
We know of no empirical estimates of the mass loading factor $\theta$
for superbubble winds.  For the larger starburst-driven galactic
winds, claims of the importance of mass-loading by evaporation of
engulfed clouds have been made based on ROSAT X-ray spectra (Suchkov
et al. 1996, della Ceca, Griffiths, \& Heckman 1997) but they are very
uncertain (Strickland \& Stevens 2000).
For a constant ambient density n$_a$, the time for the bubble to
reach a given interior density, n$_b$, will be
\begin{equation}
{\rm t_6 = 63~L_{38}^{1/4}
\left(\frac{n_b}{10^{-3}~cm^{-3}}\right)^{-3/2} n_a^{6/7}\theta^{3/2}}.
\end{equation}

We need to modify Eqn. (3) for n$_b$(t)
to account for the fact that
the density into which a bubble expands will depend on its size,
e.g., n$_a$(t) = B r$^{-p}$, where r is the radius of the region.
Statistically, the cool interstellar medium density structure can be
characterized as a fractal from $\sim$ 0.1 pc to 100 pc (Beech
1987; Bazell \& Desert 1988; Scalo 1990; Wakker 1990;  Dickman,
Horvath \& Margulis 1990; Falgarone, Phillips, \& Walker 1991;
Vogelaar, Wakker, \& Schwarz 1991; Vogelaar \& Wakker 1994) or even
to much larger (Mpc) scales (Westpfahl et al. 1999).
In three dimensions, a region of size r is likely to
contain an interior mass proportional to r$^d$, which is equivalent
to $\rho$(r) $\propto$ r$^{d-3}$.  Nearly all the above studies find
$ d \sim$ 1.3 for the two-dimensional {\it projected} density distribution,
or $p = 3 - d \sim 1.7$.
There are, however, questions concerning how these
``perimeter-area" dimensions for a projected density distribution
should be changed (if at all) for the three-dimensional distribution.
Although there is good agreement concerning the
area-perimeter dimension using various tracers, this dimension
applies to the appearance of the real three-dimensional structure
projected onto the sky.  The relation of the three-dimensional to
projected dimension is uncertain and is summarized in Westpfahl et
al. (1999).  They point out that for opaque Borel sets, the projected
dimension should be the intrinsic dimension, or 2, whichever is
smaller.  This would suggest a 3-dimensional dimension of 1.3 for the
ISM.  This value would, however, give a mass-radius relation that is much
shallower than observed (see Elmegreen \& Falgarone 1996); the
observed scaling would give a three-dimensional dimension (although
not formally the same as the perimeter-area dimension) of about 2.3.
Elmegreen \& Efremov (1999) derive a similarly large dimension from
the distribution of cloud sizes.
We are inclined to adopt the three-dimensional fractal exponent as
2.3, not 1.3, because: 1. The structures studied are not opaque but
(virtually) transparent; 2. The methods of estimating the dimension
that yield 2.3 correspond more closely with the physical basis of our
model, i.e. the number of particles or mass or average density
within a region of a certain size, rather than the perimeter-area
dimension; 3. If young stars trace out the structure of the gas from
which they formed, then the study of the manner in which the number
of star formation aggregates scales with imposed smoothing scale in
HST images of 10 galaxies by Elmegreen \& Elmegreen (2001) strongly
suggests a fractal dimension of about 2.3.  This choice of d would
give $p = 3 - d \sim 0.7$.  One should also bear in mind that
the distribution is actually multifractal (Chappell \& Scalo 2001).

For simplicity, we adopt as a fiducial scaling relation Larson's (1981)
scaling relation for molecular clouds,
n$_a$(r) $\sim$ 10$^3$~B$_3$~cm$^{-3}$r$_{pc}^{-1}$,
where B$_3 \sim 1.7$ is a normalization constant
representing the number density in units of 10$^3$ cm$^{-3}$
at the scale of one parsec.
We realize that this relation may be seriously affected by
selection effects (Kegel 1989; Scalo 1990).  We also examine the
cases for $p = 0.7, 1.7$ to check the sensitivity.  For the case
$p = 1$ we obtain:
\begin{equation}
{\rm n_b(t) = 0.1~cm^{-3}~B_3^{5/7} L_{38}^{1/42} t_6^{-23/21}\theta }.
\end{equation}
The interior bubble density decreases more rapidly with time than in the
case for constant n$_a$ because the interior volume is increasing more
rapidly with time.  For comparison, if mass loading dominates but
with a constant mass injection rate ${\dot M}$, n$_b$(t)$\propto$
t$^{-5/4}$,
while if there is no mass loading, and n$_b$ is governed completely
by expansion, n$_b$(t)$\propto$ t$^{-9/4}$.  If we took
p = 0.7 (our preferred value), n$_b$ would be larger by about a factor
of 2.  If we took p = 1.7, n$_b$ would be smaller by about a factor of 7.
        
Solving Eqn. (5)
for the time to reach internal density n$_b$, we get
\begin{equation}
{\rm t_6 = 66~B_3^{15/23} L_{38}^{1/46}
\left(\frac{n_{b,3}}{\theta}\right)^{-21/23}}.
\end{equation}
This time would be longer by a factor of about 5 if $p = 0.7$
and shorter by a factor of about 12 for $p = 1.7$.
Any constraints on the lifetime of the progenitor star thus depend rather
sensitively on the structure of the ambient medium.  We will return to
this topic on \S 4. For now we conclude that afterglow density
estimates less than about 0.1 cm$^{-3}$ are consistent with
$\gamma$-ray bursts exploding into pre-existing superbubbles,
independent of the specific mechanism of the $\gamma$-ray
bursts. In addition, the factor $\theta$ expressing the
uncertainty in the mass loading rate was estimated to be in the
range 0.1 to 3.  With hydrodynamical mass loading and no
conduction (Strickland \& Stevens 1998), the bubble density
will be smaller by an order of magnitude and the corresponding
time larger by about the same factor.

\section{VARIATIONS IN AFTERGLOW DENSITIES}

    There are a number of effects that will provide variations in the
density into which a $\gamma$-ray burst might explode within
the context of this hypothesis that $\gamma$-ray bursts
propagate into superbubbles.  Each of these has potentially
different implications for the progenitors of $\gamma$-ray
bursts.

    1. Even if the $\gamma$-ray burst explodes within the
cluster that produced the superbubble, so that the $\gamma$-ray
burst is roughly centrally located in the bubble, there are
bound to be variations in the in the ambient density n$_a$. For
example, if the wind initially expands within a giant molecular
cloud (GMC), the mean density may be 100 cm$^{-3}$, but there
will be variations in the mean value from cloud to cloud, and
GMC internal density fluctuations of several orders of
magnitude. Most of these internal cloud density fluctuations
will be on scales smaller than the bubble size at later times. 
The effects of superbubbles that begin their expansion at
different distances from the midplane of a galactic disk
(Silich et al. 1994, 1996) will introduce further variations in
ambient density. Note that even large variations in the cluster
wind kinetic energy L$_{38}$, reflecting different masses of
clusters, will not substantially affect our results (cf. eqn.
5).

    2. Superbubbles have a variety of ages, and hence interior densities
into which $\gamma$-ray bursts may explode.  If the clusters
giving rise to the superbubbles are born with a rate that is a function of
time given by B$_\gamma$(t), then the probability distribution of
superbubbles with interior density n$_b$ is given by
            \begin{eqnarray}
{\rm f(n_b) =\frac {B_\gamma[t(n_b)]}{|dn_b/dt|}}.
\end{eqnarray}
Assuming a constant rate of cluster formation and taking ${\rm |dn_b/dt|}$
from
Eqn.(5) gives
${\rm f(n_b) \propto n_b^{-44/23}}$ showing
that we are much more likely to observe a $\gamma$-ray burst
exploding into a superbubble of low density, basically
because the superbubbles decelerate with time, so
more shells occur at large ages and small densities.
It can be shown using the relations given above that in the limits of
expansion or mass loading dominance, the
exponent of f(n$_b$) would be $\sim$ -3/2 to -2, nearly independent of the
dependence
of R on n$_a$.  Thus the conclusion that the inferred
afterglow densities would be dominated by the smallest values of
n$_b$ in the absence of other effects is robust with respect to
assumptions about the mass loading.  The $\gamma$-ray bursts
may not, of course, explode randomly, but may be correlated in
time and space with a given superbubble.

    3. Given the sizes of superbubbles, $\sim 10$~pc to 1 kpc, it is likely
that a given superbubble has engulfed another, younger,
cluster.  In this case, a $\gamma$-ray burst exploding in an
engulfed cluster will expand somewhere within the earlier
superbubble (and within the ambient medium of its host
cluster).  The conduction solution of Weaver et al. (1997) and
the no-conduction simulations
of Strickland \& Stevens (1998) both exhibit radial
density profiles with significant variations, from one to two orders of
magnitude.  Thus, although the $\gamma$-ray burst is most
likely to explode in the ``plateau" region of the density distribution
(basically given by Eqn. (1)),
there is a significant probability
that it will explode in smaller or larger densities.  We point out that
Chu
\& MacLow (1990) proposed that supernova remnants
explode off-center in superbubbles in order to explain the X-ray emission of
H II complexes in the LMC.
 
    4. Considering the collimated nature of the $\gamma$-ray
burst explosion in the model of Panaitescu \& Kumar (2001) and
suspected in general (e.g. Frail et al. 2001), the shock has a
probability of encountering one of the clouds engulfed by the
superbubble, or one of the many supernova blast waves that
impose sizeable density fluctuations within the superbubble.

    Given all these considerations, we conclude that it is likely that the
inferred ambient densities for $\gamma$-ray burst afterglows
could span a range of four or five orders of magnitude, as inferred
empirically by Panaitescu \& Kumar (2001), and we can easily
explain both the lowest and higher inferred ambient densities.

    A consequence of our proposal that $\gamma$-ray bursts
explode in pre-existing superbubbles is that, at high
resolution, $\gamma$-ray bursts with low ambient densities
should be spatially associated not with the bluest regions of
galaxies, but with X-ray bright spots associated with
superbubbles.  Perhaps this explains why Holland et al.  (2001)
find that GRB 980703, one of the low density cases, shows no
connection with any special features of the host. With a
resolution of about 1/2 arcsec, Chandra X-ray observations
would only be able to resolve medium size, 100 pc, superbubbles
at distances less than $\sim$ 20 Mpc. The best H I 21 cm
interferometer mappings of holes in galactic gas can reach
somewhat larger distances.  This resolution limit corresponds
to a redshift of about z = 0.003.  Since the mean redshift of
$\gamma$-ray bursts is $\ga$ 1, the probability of finding such
a nearby
$\gamma$-ray burst is $\la 3\times 10^{-8}$ per event.  

\section{CONSTRAINTS ON PROGENITORS}

The evolution of superbubbles is potentially complex and so
evaluating the implication of $\gamma$-ray bursts occuring in
superbubble environments is uncertain.  Here we will survey
some of the reasonable possibilities.

One possibility is that $\gamma$-ray bursts occur in some type
of coalescing binary, e.g. neutron stars.  Such a possibility
requires rather short-lived binaries since all identified
$\gamma$-ray bursts so far are within the optical contours of
the host galaxy (Fruchter, private communication) and hence
cannot have drifted very far before coalescence. Such a model
might be consistent with both an overall correlation with star
formation but with a lack of universal correlation with
specific blue knots of recent star formation.  Drifting binary
neutron stars might be expected to randomly sample the complex
ISM expected in a star forming galaxy that blows bubbles, as
outlined in \S 3.

In the remainder of this section, we will consider possible
constraints on massive stars as the progenitors of $\gamma$-ray
bursts. We will consider constant power input to the bubbles,
but according to the models of Shull \& Saken (1995), varying
the power input, e.g. from continuous to coeval star
formation will not change any of these conclusions substantially.

The simplest hypothesis is that there is a coeval burst of star
formation in a cluster after which the stars themselves blow winds
to make the bubble
and eventually die as supernovae.  This hypothesis is especially
interesting because it implies that if the $\gamma$-ray bursts
that go off in the low density environments are, in fact,
within such self-generated bubbles (cf. points 1 and 2 of \S
3), then the stars that produce the $\gamma$-ray bursts are not
the most massive stars. Some stars must already have evolved
with strong winds and perhaps died to blow a sufficiently low
density bubble.  This raises the possibility of placing an {\it
upper limit} on the progenitor mass of $\gamma$-ray burst
progenitor stars.

Eq (4) applies to the simple case of coeval evolution
of the stars and expansion of the bubble into a constant density
environment.  The implication is that, for a
bubble driven with approximately constant power, the ambient density
into which the bubble propagates must
be very low, ${\rm n_a} \la 0.05$ cm$^{-3}$ to
allow time, about 5 million years,
for, say, 30 M$_\odot$ stars to evolve and explode in a bubble
of mean interior density of  n$_b = 10^{-3}$ cm$^{-3}$.
Such a low value of ${\rm n_a}$ suggests a pre-existing bubble,
a case we consider below.
If the mass loading of the bubble is hydrodynamical rather than
by conduction, the ambient density for the $\gamma$-ray burst
can be significantly larger; however is the mass loading is due
to evaporation of engulfed clouds, the required ambient
densities must be somewhat smaller.
 If the ambient density is higher, then even lower mass stars must
have had time to evolve before the first $\gamma$-ray burst
went off at such low bubble densities.

The possibility of a $\gamma$-ray burst progenitor of mass
$\la$ 30 M$_\odot$ in order to give time for a coeval starburst
to form a low density bubble is strongly constrained
by considering the rate of occurence of $\gamma$-ray bursts.
Scalo \& Wheeler (2001) estimate that the ratio of $\gamma$-ray
bursts to supernovae is about one in several thousand if
collimation is neglected. Even with rather strong collimation
into one part in 100 of 4$\pi$ steradians, the lowest mass that
could contribute to $\gamma$-ray bursts would be over 100
M$_\odot$.  Only if the collimation of $\gamma$-ray bursts were
substantially smaller could the rate of $\gamma$-ray bursts be
comparable to the rate of death of stars of 30 M$_\odot$ or
less. This seems extreme, but we note that models for the
afterglow imply that some bursts are collimated to this degree
(Panaitescu \& Kumar 2001; Frail et al. 2001). Other ways to
avoid excessively large
$\gamma$-ray burst rates with the low mass progenitors demanded
in the coeval bubble picture are that $\gamma$-ray bursts do
not arise from stars with an upper limit threshhold mass, but
occur in a narrow mass range or from a small
fraction of events with some special extreme of character, e.g.
rotation or magnetic field, over a broad range of masses.

This picture is modified somewhat if we consider bubbles expanding
into ``fractal" density distribution, as described by Eqn
(6).
For ${\rm B_3 \sim 1}$ and ${\rm p \la 1}$, the time to reach
low densities, ${\rm n_b \sim 10^{-3}}$ cm$^{-3}$, is very long
so that unrealistically small progenitor masses would be required.
For p = 1.7, however, the time to reach these small bubble densities
is fairly short.  The upper limit to the progenitor mass of a
$\gamma$-ray burst might be consistent with the estimated rates
of $\gamma$-ray bursts and still allow time for even higher
mass stars to blow the requisite bubble. The upper limit would,
clearly, be a rather sensitive function of the parameter p
describing the distribution of ambient gas. There is additional
uncertainty due to the mass loading parameter, $\theta$. Another
interesting alternative is explored by Shull \& Saken (1995).
They investigate the scenario proposed by Doom et al. (1985)
wherein lower mass stars, say about 15 M$_\odot$ are born first
and the most massive stars are only born later, after an
interval of 10 to 20 million years.  In this case, the older,
lower mass stars blow the bubble, but the younger, higher mass
stars could provide the $\gamma$-ray bursts.  This possibility
obviously precludes determining an upper limit to $\gamma$-ray
burst progenitors.  Rather, it might be possible to constrain
the lower limit, but this would depend on the parameter, p, of
the ambient density structure, the mass loading parameter,
$\theta$, and the time history of the SFR.

The inevitability of bubbles in regions of active star formation
leads to the possibility that a $\gamma$-ray burst will explode
in a cluster that has itself been engulfed by a older,
independent superbubble, cf. point 3 of \S 3.  In this
case, an older cluster could have blown a bubble and then
a younger cluster, perhaps formed by the compression of the
shell of the first one could produce the $\gamma$-ray burst. 
In this case it is difficult to put any constraints at all on
the progenitor of the $\gamma$-ray burst.  One problem with
this possibility is that the remnant density in the younger
star cluster might be larger than can be tolerated for the
lowest densities revealed by the afterglows.

\section{Conclusions}

The low densities in which some $\gamma$-ray burst afterglows
propagate provide interesting clues to the environment of
$\gamma$-ray bursts and to their progenitors.  Our principle
conclusion is that superbubbles can easily provide such
environments. In general, to attain low densities $\sim
10^{-3}$ cm$^{-3}$, the superbubbles must propagate into
relatively low ambient densities or must be rather old.  The
expected evolution of superbubbles favors large, low density
bubbles.

The low afterglow densities could be consistent with either the
hypothesis that rather young and slowly drifting neutron
star binaries randomly sample the large expected density
variation of active star forming galaxies or with a variety of
possibilities associated with massive star progenitors.  The
low ambient densities for some afterglows do not {\it a priori}
preclude massive star progenitors for $\gamma$-ray bursts.

The expected superbubble properties of star-forming galaxies can,
in principle, constrain the progenitor masses if
$\gamma$-ray bursts arise in massive stars, but in practice
uncertainties in ISM structure, bubble mass loading, SFR
history, cluster evolution, and stellar mass functions, make it
difficult to do so quantitatively.

While the interior of superbubbles provides a natural environment
for low ambient afterglow densities, the self-contamination of such
a low density environment by a stellar wind remains a severe
problem for the massive star hypothesis.
To be compatible with the lack of any evidence for such
a wind, the wind density must be very low.  This
sets constraints on either the mass loss rate or the wind
velocity or both.  For a stellar wind characterized by a constant velocity
wind at $10^8~{\rm v_8}$~cm~s$^{-1}$ carrying
mass at a rate $10^{-5}{\dot M_{-5}}$~M$_\odot$~yr$^{-1}$,
the baryon number density is
\begin{equation}
{\rm n = 30~{\rm cm^{-3}}~{\dot M_{-5}}v_8^{-1}R_{17}^{-2}},
\end{equation}
ignoring whether the baryons are single or incorporated in
nuclei, and with a radius, R$_{17}$, in units of $10^{17}$ cm,
a characteristic radius to which afterglows propagate.  Such a wind,
as might characterize a typical O or Wolf-Rayet star, is incompatible with
the lowest afterglow ambient densities inferred.  For the density in
such a wind to be less than the lowest ambient
densities $\sim 10^{-3}$ cm$^{-3}$ at a radius
$\sim 10^{17}$ cm, one requires
${\rm \dot M_{-5} v_8^{-1}}$ to be less than $\sim 10^{-4}$.  This
is a rather extreme requirement, but it may be fulfilled by
stripped cores with fast winds and atmospheres dominated by
heavy elements, e.g. carbon and oxygen, that are difficult
to expel by radiation pressure due to their large weight.
On the other hand heavy ions will have more lines to interact
via radiative acceleration, so it is not clear that the winds can be
suppressed.   In any case, this suggests that for the wind of a massive star
to not affect the afterglow, something like
a Type Ic supernova makes a natural progenitor, a point made in other
contexts (Woosley 1993; MacFadyen \& Woosley 1999; Wheeler
et al. 2000).  
Another possibility is that the mean wind density is high, but that
a small column, for instance along the rotation axis, has a much lower
density.  The nature of the wind must be addressed to reconcile
the low ambient afterglow densities with massive star progenitors.

Note that the column depth in a wind is
\begin{equation}
l = 50~{\rm g~cm^{-2}}~{\dot M_{-5}}v_8^{-1}R_{10}^{-1},
\end{equation}
where the radius is in units of $10^{10}$ cm, characteristic
of the outer radius of the core of a massive star.
For {\rm $\dot M_{-5}$v$_8^{-1}\sim 1$},
the column depth is high enough to suppress $\gamma$-ray bursts
emitted at R $\sim 10^{10}$ cm.
If $\dot M_{-5}$v$_8^{-1} \lesssim 10^{-4}$ in order to
provide low densities at large distances, then there
will also be negligible column depth in the wind.

We predict that the $\gamma$-ray bursts with low ambient
densities will be identified with X-ray bright regions of
galaxies and HI holes, corresponding to superbubbles, rather
than with blue localized regions of star formation.

\begin{center}
Acknowledgements
\end{center}

We are grateful to Andy Fruchter, Peter H\"oflich, Pawan Kumar and
Brad Schaefer for discussions of $\gamma$-ray bursts and their
afterglows and to Doug Swartz for discussions of superbubbles. 
We especially thank Sally Oey for guiding us to relevant
literature on superbubble interiors that formed the basis for
this work.  This research was supported by NSF grants
AST-9818960, AST-9907528, and NASA Grant HST-GO-08688.09.

\end{document}